\documentclass[pre,twocolumn,aps]{revtex4}
\usepackage{graphicx}
\usepackage{comment}
\usepackage{bm}
\usepackage{amsmath,amssymb} 
\usepackage{epstopdf}
\usepackage{verbatim}
\usepackage{float}
\newcommand{\bR}{{\bf r}_1, ..., {\bf r}_N}

\begin{document}

\title{Do all liquids become strongly correlating at high pressure?}
\author{Jon J. Papini, Thomas B. Schr{\o}der, and Jeppe C. Dyre}
\email{dyre@ruc.dk}
\affiliation{DNRF Centre ``Glass and Time'', IMFUFA, Department of Sciences, Roskilde University, Postbox 260, DK-4000 Roskilde, Denmark}
\date{\today}

\begin{abstract}
We present molecular dynamics simulations studying the influence of pressure on the correlation between the constant-volume thermal equilibrium fluctuations of virial $W$ and potential energy $U$, focusing on liquids that are not strongly correlating at low pressure, i.e., do not have a $WU$ correlation coefficient above 0.9. The systems studied are the two hydrogen-bonded liquids GROMOS methanol and TIT5P water, the ionic liquid defined by a united-atom model of the 1-butyl-3-methyl-imidazolium nitrate and, for reference, the standard single-component Lennard-Jones liquid. The simulations were performed for pressures varying from 0 GPa to 10 GPa. For all systems studied we find that the virial / potential energy correlation increases with increasing pressure. This suggests that if crystallization is avoided, all liquids become strongly correlating at sufficiently high pressure.
\end{abstract}
\pacs{??}

\maketitle

The properties of strongly correlating liquids were recently discussed in several papers \cite{scl}. These liquids by definition exhibit strong correlations between their constant-volume equilibrium fluctuations of the potential energy $U$ and the virial \cite{all87,han05} $W\equiv -1/3 \sum_i {\bf r}_i \cdot {\bf \nabla}_{{\bf r}_i} U(\bR)$, where ${{\bf r}_i}$ is the position of particle $i$. Recall that, if $p$ is the pressure, $V$ the volume, $N$ the number of particles, and $T$ the temperature, the average virial $\langle W\rangle$ gives the configurational contribution to the pressure \cite{all87,han05}:

\begin{equation}\label{Wdef}
pV \,=\,
Nk_BT+\langle W\rangle\,.
\end{equation}
If $\Delta$ denotes the instantaneous deviations from equilibrium mean values, the $WU$ correlation is quantified by the correlation coefficient $R$ defined by (where angular brackets denote $NVT$ ensemble averages, i.e., averages at constant volume and temperature)

\begin{equation}\label{R}
R \,=\,
\frac{\langle\Delta W\Delta U\rangle}
{\sqrt{\langle(\Delta W)^2\rangle\langle(\Delta U)^2\rangle}}\,.
\end{equation}
Perfect correlation gives $R=1$ and strongly correlating liquids are defined \cite{scl} by $R\ge 0.9$.

The computer simulations of Ref. \onlinecite{scl} indicate that the correlation coefficient $R$ tends to increase at increasing pressure, but no systematic studies have been carried out of the effect of pressure on the correlation. The simulations of Ref. \onlinecite{scl} showed that van der Waals type liquids and metallic liquids are generally strongly correlating. In contrast, liquids composed of molecules whose interactions have competing or directional interactions are generally not strongly correlating. The latter classes of liquids include the hydrogen-bonded liquids, the covalently bonded liquids, and the strongly ionic liquids. Since previous works indicated that $R$ increases at increasing pressure, one may ask whether all liquids become strongly correlating at sufficiently large pressure. This is the question addressed in the present brief report.

Simulations of four different model liquids were performed with $NVT$ molecular dynamics using the Gromacs package \cite{gromacs}. For each model samples of different densities were created and mixed during 1 ps (argon units) at a high temperature, followed by a ramping down to the desired isotherm. Here the systems were equilibrated at constant temperature during 10 - 500 ns. The data for each state point shown below represent an average taken over five statistically independent samples, with a sampling frequency of 0.2 ps and production runs of length 10 ns. The following systems were studied: 1) {\it The single-component Lennard-Jones liquid} defined by the pair potential $v_{LJ}(r) = 4 \epsilon \left[(\sigma /r)^{12}-(\sigma /r)^{6}\right]$. This system serves as a reference strongly correlating liquid. The results reported below refer to standard argon units ($\sigma = 0.34$ nm, $\epsilon = 0.997$ kJ/mol). Samples consisting of $N=864$ particles were studied. 2) {\it Methanol}: The GROMOS force field was used \cite{Gromos,Methanol}, which is composed of three sites representing, respectively, the methyl group, the oxygen atom, and the oxygen-bonded hydrogen atom (H). The masses are, respectively, 15.035 u, 15.999 u, 1.008 u; the Coulomb interactions are given by the following charges: 0.176 e, -0.574 e, and 0.398 e. The sites interact with sites on other methanol molecules by additional Lennard-Jones interactions with the constants $\epsilon_{MM}=0.9444$ kJ/mol, $\epsilon_{OO} = 0.8496$ kJ/mol, $\epsilon_{MO} = 0.9770$ kJ/mol, $\sigma_{MM} = 0.3646$ nm, $\sigma_{OO} = 0.2955$ nm and $\sigma_{MO} = 0.3235$ nm. The van der Waals interactions are cut off smoothly between $0.9$ nm and $1.1$ nm. The M-O distance is fixed at $0.136$ nm, the O-H distance at $0.1$ nm, and the M-O-H bond angle at $108.53^o$. Samples consisting of $N=1728$ molecules were studied. 3) {\it TIP5P}: In this water model \cite{Tip5p} each water molecule is described by five sites: one site represents the oxygen atom (O), two sites represent the hydrogen atoms, and two sites locate the centers of negative charge that correspond to the oxygen lone-pair electrons. The potential parameters and charges used are the same as in Ref. \onlinecite{scl}. Sample consisting of $N=512$ molecules were studied. 4) {\it [BMIM]$^{+}$[NO$_{3}$]$^{-}$}: A united-atom model of the ionic liquid 1-butyl-3-methyl-imidazolium nitrate \cite{Imadno3} based on the GROMOS \cite{Gromos} force field. The same parameters were used as in Ref. \onlinecite{Imadno3}.

\begin{figure}[ht]
\includegraphics[width=0.9\linewidth,clip=true]{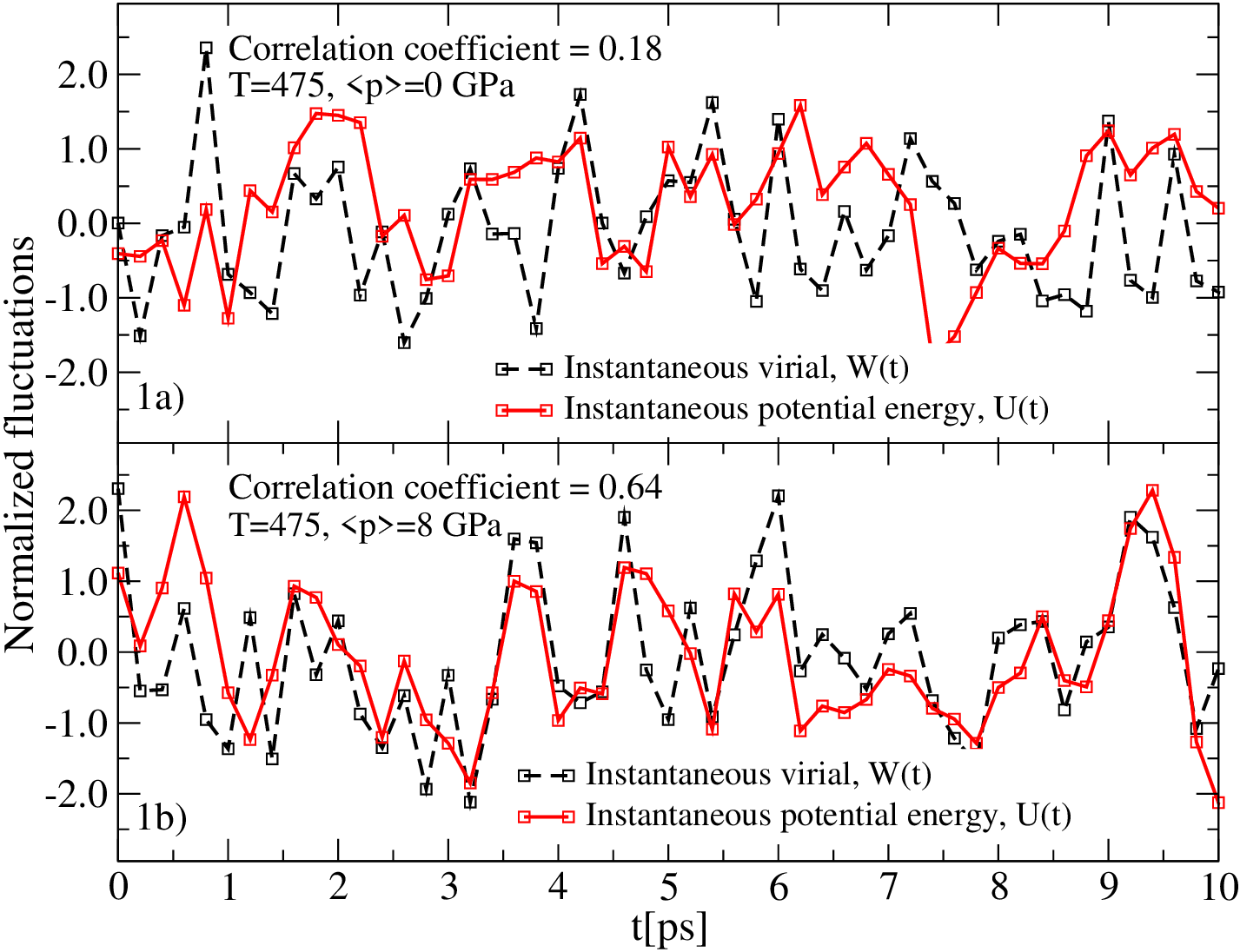}
\caption{Time series from simulation of 512 molecules of TIP5P water in the NVT ensemble at two different pressures. (a) At zero pressure the correlation between normalized fluctuations of the virial, $\triangle W(t)/\sqrt{\langle (\triangle W)^2\rangle}$ and that of the potential energy, $\triangle U(t)/\sqrt{\langle (\triangle U)^2\rangle}$, is weak, with a correlation coefficient of $R=0.18$. As shown in Ref. \onlinecite{scl} this low correlation is related to the existence of a density maximum at lower temperature where the $WU$ correlation is almost zero. (b) At the pressure 8 GPa the $WU$ correlation is considerably stronger ($R=0.64$).}\label{Figure1}
\end{figure}

As an example of $WU$ correlations Fig. \ref{Figure1} shows the equilibrium fluctuations as a function of time of the TIP5P water model's normalized virial and potential energy. Figure \ref{Figure1}(a) gives data from a simulation at zero pressure at $T=475$ K. The fluctuations of $W$ and $U$ are rather uncorrelated ($R=0.18$). At lower temperatures the correlation is even lower; indeed near the density maximum the correlation is close to zero \cite{scl}. Figure \ref{Figure1}(b) gives data from a simulation at $p=8$ GPa at the same temperature. Here the correlation is significantly larger ($R=0.64$).

\begin{figure}
\includegraphics[width=0.9\linewidth]{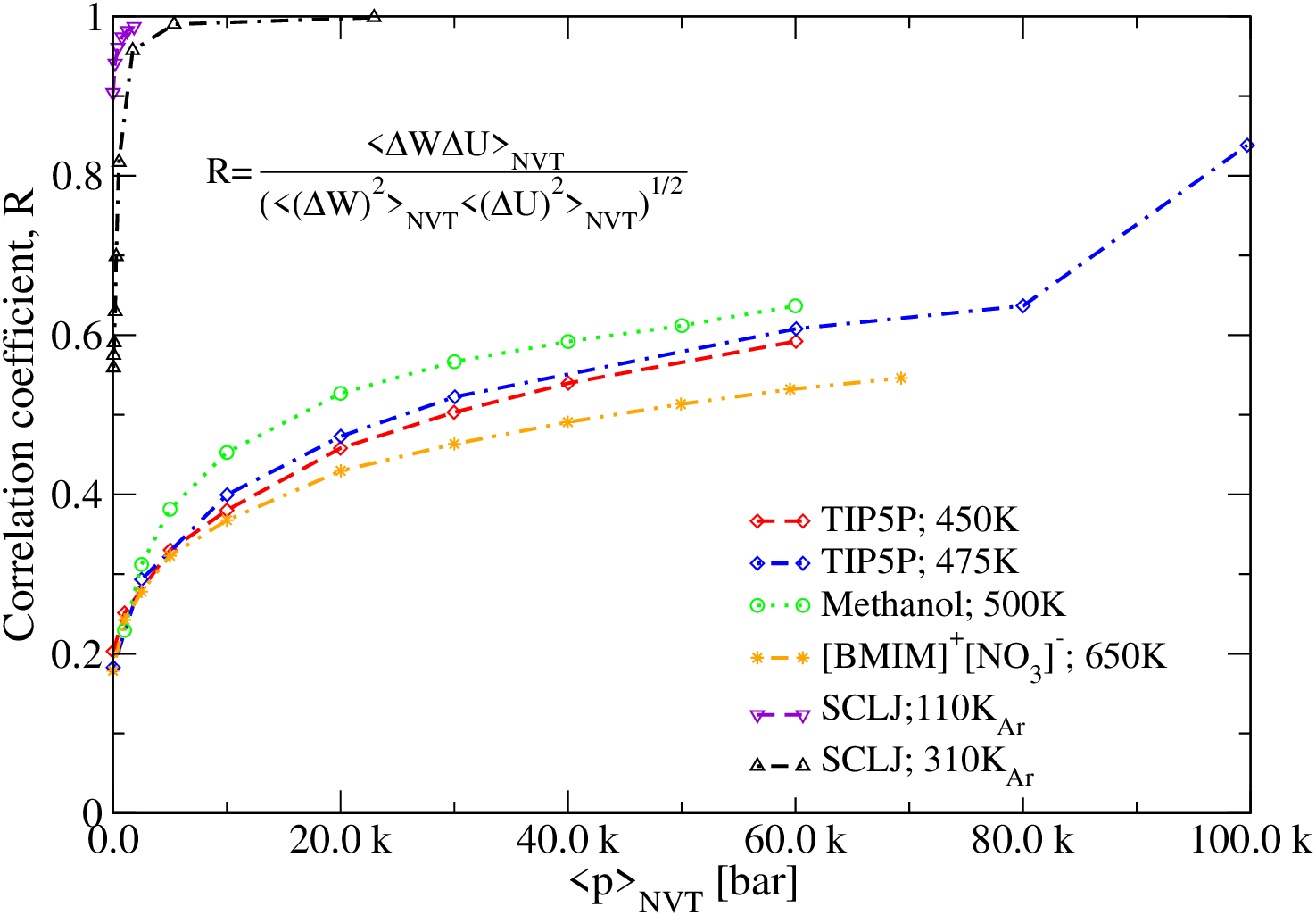}
\caption{The $WU$ correlation coefficient $R$ plotted as a function of pressure along isotherms for the following systems: 1) The standard, single-component Lennard-Jones liquid (two isotherms), 2) the ionic liquid [BMIM]$^{+}$[NO$_{3}$]$^{-}$, 3) methanol, and 4) the TIP5P water model (two isotherms; the last point represents a crystallized sample). In all cases the correlation increases with increasing pressure. Data were taken from 10 ns of simulations of each liquid.}\label{Figure2a}
\end{figure}

To systematically investigate the influence of pressure on the $WU$ correlation we calculated the correlation coefficient as a function of pressure along isotherms for the four liquids. Figure \ref{Figure2a} shows that the correlation increases with increasing pressure for all systems. The rather low correlation in the case of the single component Lennard-Jones (SCLJ) liquid at T=310 K reflects the fact that only the last three points stem from liquid-state simulations.

\begin{figure}[ht]
\includegraphics[width=0.9\linewidth,clip=true]{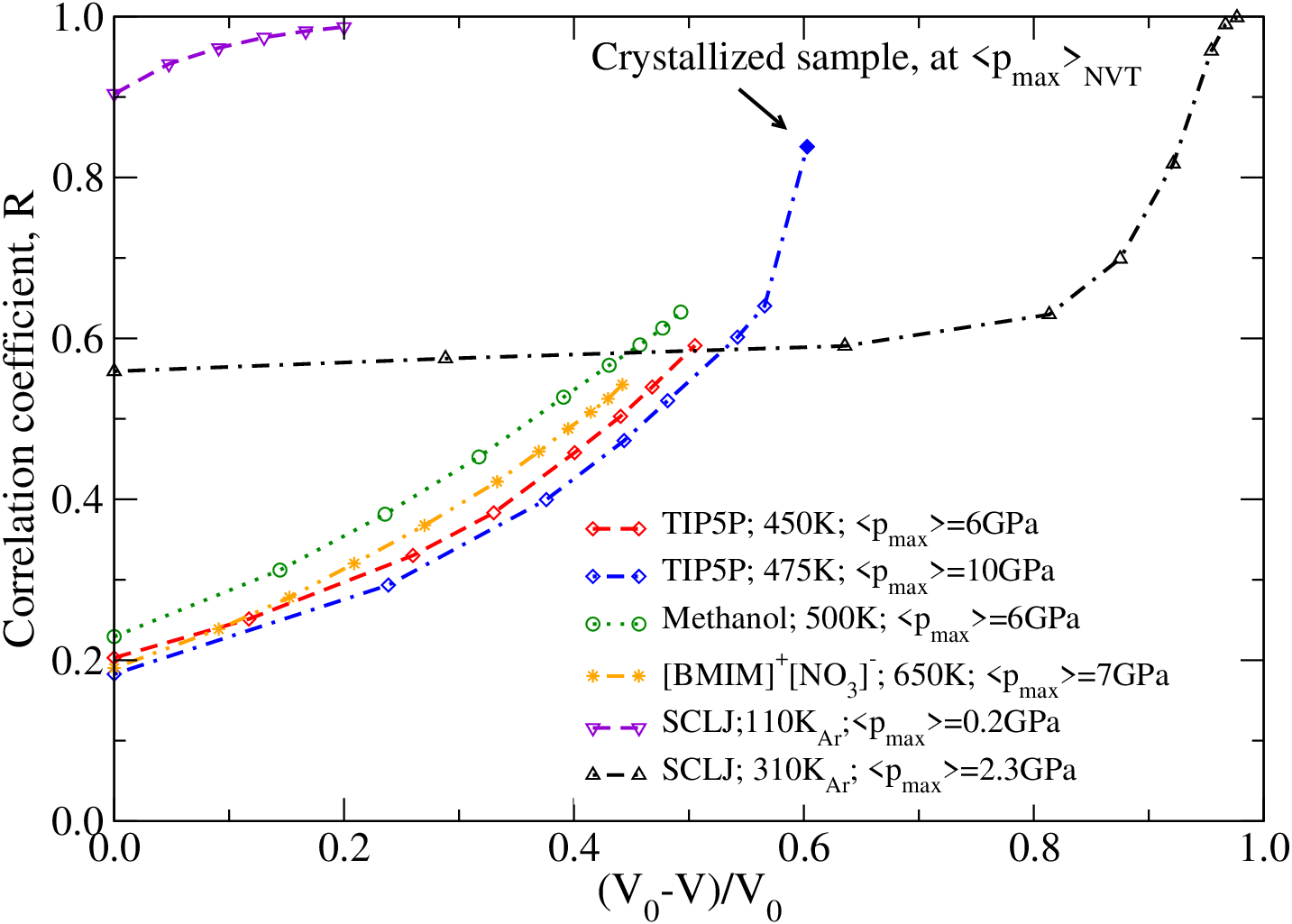}
\caption{The correlation coefficient $R$ plotted against the volume decrease relative to the volume $V_0$ at the lowest pressure of the given isotherm. The higher density isotherm (450 K) of TIP5P water shows stronger correlation than its less dense counterpart (475 K) at the same relative volume change.}\label{Figure2b}
\end{figure}

In Fig. \ref{Figure2b} the correlation coefficient was plotted instead as a function of the relative volume change, $\triangle V= (V_0-V)/V_0$, where $V_0$ is the highest volume at the given temperature corresponding to the lowest pressure of the simulation. In three cases the lowest pressure was around $1$ bar, i.e., effectively close to zero, but for Methanol the lowest pressure was of order $0.1$ GPa. Water crystallizes upon compression before it reaches the correlation coefficient $R>0.9$ that defines a strongly correlating liquid. 

In summary, all liquids studied show increasing virial / potential energy correlations as pressure increases. These simulations indicate that if crystallization is avoided, all liquids become strongly correlating at sufficiently high pressure.

\section{Acknowledgments}

Nuno Micaelo is gratefully acknowledged for providing the coordinates of an equilibrated sample of the united atom model of 1-butyl-3-methyl-imidazolium nitrate. 
The center for viscous liquid dynamics ``Glass and Time'' is sponsored by The Danish National Research Foundation.

\end{document}